\documentclass[aps,prl,twocolumn,showpacs,superscriptaddress]{revtex4}
\usepackage{amsmath}
\usepackage{dcolumn}
\usepackage{epsfig}
\usepackage{graphicx}
\usepackage{latexsym}

\def\YBCO{YBa$_2$Cu$_3$O$_{6+x}$}

\begin{document}
\hyphenation{Ka-pi-tul-nik}



\title{Polar Kerr Effect Measurements of \YBCO: Evidence for  Broken Symmetry Near the Pseudogap Temperature}

\author{Jing Xia}
\affiliation{Department of Physics, Stanford University, Stanford, CA 94305}
\affiliation{Geballe Laboratory for Advanced Materials, Stanford University, Stanford, California, 94305}
\author{ Elizabeth Schemm}
\affiliation{Department of Physics, Stanford University, Stanford, CA 94305}
\affiliation{Geballe Laboratory for Advanced Materials, Stanford University, Stanford, California, 94305}
\author{G. Deutscher}
\affiliation{School of Physics and Astronomy, Tel Aviv University, Tel Aviv 69978, Israel}
\author{S.A. Kivelson}
\affiliation{Department of Physics, Stanford University, Stanford, CA 94305}
\affiliation{Geballe Laboratory for Advanced Materials, Stanford University, Stanford, California, 94305}
\author{D. A. Bonn}
\affiliation{Department of Physics, University of British Columbia, Vancouver, B.C., V6T2E7, Canada }
\author{W. N. Hardy}
\affiliation{Department of Physics, University of British Columbia, Vancouver, B.C., V6T2E7, Canada }
\author{ R. Liang}
\affiliation{Department of Physics, University of British Columbia, Vancouver, B.C., V6T2E7, Canada }
\author{W. Siemons}
\affiliation{Geballe Laboratory for Advanced Materials, Stanford University, Stanford, California, 94305}
\affiliation{MESA+ Institute for Nanotechnology, Twente University, Enschede 7500 AE, The Netherlands}
\author{G. Koster}
\affiliation{Geballe Laboratory for Advanced Materials, Stanford University, Stanford, California, 94305}
\affiliation{MESA+ Institute for Nanotechnology, Twente University, Enschede 7500 AE, The Netherlands}
\author{ M. M. Fejer } 
\affiliation{Department of Applied Physics, Stanford University, Stanford, CA 94305}
\author{A. Kapitulnik}
\affiliation{Department of Physics, Stanford University, Stanford, CA 94305}
\affiliation{Geballe Laboratory for Advanced Materials, Stanford University, Stanford, California, 94305}
\affiliation{Department of Applied Physics, Stanford University, Stanford, CA 94305}

\date{\today}

\begin{abstract}
Polar Kerr effect in the high-Tc superconductor \YBCO~was measured at zero magnetic field with high precision using a cyogenic Sagnac fiber interferometer. We observed non-zero Kerr rotations of order $\sim 1 \mu$rad appearing near the pseudogap temperature $T^*$, and marking what appears to be a true phase transition. Anomalous magnetic behavior in magnetic-field training of the effect suggests that  time reversal symmetry is already broken above room temperature.
\end{abstract}


\maketitle

One of the most challenging puzzles that has emerged within the phenomenology of the high-temperature superconductors (HTSC) is to understand the occurrence and role of the normal-state ``pseudogap" phase in underdoped cuprates  \cite{pseudoreview}.  This phase exhibits anomalous behavior of many properties including magnetic \cite{alloul}, transport \cite{ito}, thermodynamic \cite{loram}, and optical properties below a temperature, $T^*$, large compared to the superconducting (SC) transition temperature, $T_c$.  Two major classes of theories have been introduced in an attempt to describe the pseudogap state: One in which the pseudogap temperature $T^*$ represents a crossover into a state with preformed pairs with a d-wave gap symmetry \cite{lee1,ek}, and  another in which  $T^*$ marks a true transition into a phase with broken symmetry which ends at a quantum critical point, typically inside the superconducting dome. While at low-doping this phase may compete with superconductvity, it might provide fluctuations that are responsible for the enhanced transition temperature near its quantum critical point (e.g. as in ref.~\cite{varma}). Examples include competing phases of charge and spin density waves \cite{sketal}, or charge current loops which either do  \cite{sudip} or do not \cite{varma,simon} break translational symmetry. 

\begin{figure}[h]
\begin{center}
\includegraphics[width=1.0 \columnwidth]{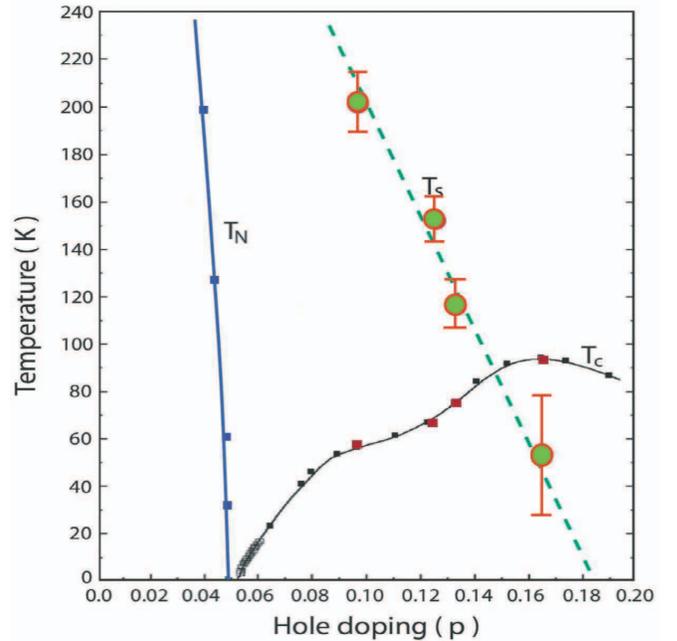}
\end{center}
\caption{ Onset of Kerr effect signal, $T_s$ (circles), and $T_c$ (red squares) for the  \YBCO~samples reported in this paper.  Also shown are $T_c(p)$ (from \cite{liang}), and $T_N(p) $ (from  \cite{lavrov}). } 
\label{pd}
\end{figure}

In this paper, we report high resolution optical Kerr effect measurements on \YBCO~crystals with various hole concentrations $p$.  ($p$ is, in turn, a monotonic function of the oxygen concentration, $x$, and it also depends on oxygen ordering in the chains \cite{liang}.)  We identify a sharp phase transition at a temperature $T_s(p)$, below which there is a non-zero Kerr angle, indicating the existence of a phase with broken time reversal symmetry (TRS). Both the magnitude and hole concentration dependence of $T_s$ are in close correspondence with those of the pseudo-gap crossover temperature, $T^*$, which has been identified in other physical quantities. In particular, as shown in Fig.~\ref{pd}, $T_s$ is substantially larger than the superconducting $T_c$ in underdoped materials, but drops rapidly with increasing hole concentration, so that it is smaller than $T_c$ in a near optimally doped crystal and extrapolates to zero at a putative quantum critical point under the superconducting dome. The magnitude of the Kerr rotation in \YBCO~is smaller by $\sim$4 orders of magnitude than that observed in other itinerant ferromagnetic oxides \cite{lsmomo,sromo}, and the temperature dependence is ``superlinear" near $T_c$, suggesting that we are either not directly measuring the principal order parameter which characterizes the pseudo-gap phase in YBCO, or we measure its very small ``ferromagnetic-like" component. In addition we find a hysteretic memory effect which seemingly implies that TRS is broken in all cases at a still higher temperature (above room temperature), although no Kerr effect is detectable within our sensitivity at temperatures above $T_s$.

High quality \YBCO~single crystals with $x=0.5$ (ortho-II, $T_c=59$K), $x=0.67$ (ortho-VIII, $T_c=65$K), $x=0.75$ (ortho-III, $T_c=75$K), and $x=0.92$ (ortho-I, $T_c=92$K), were grown by a flux method in BaZrO$_3$ crucibles \cite{liang}. The crystals, in the form of ($ab$-plane) platelets several millimeters on a side and a fraction of a millimeter thick ($c$-direction), were mechanically detwinned. X-ray diffraction measurements indicate long uninterrupted chain lengths (e.g. for ortho-II this length is $\sim 120\times$b \cite{liang}).  Polar Kerr effect measurements were performed using a  zero-area-loop polarization-Sagnac interferometer at wavelength of $\lambda =$1550 nm  \cite{xia2}.  The same apparatus was previously used to detect (a noticeably weaker) TRS-breaking below $T_c$ in Sr$_2$RuO$_4$ \cite{xia1}. Typical performance was a shot-noise limited 0.1 $\mu$rad/$\sqrt{Hz}$ at 10 $\mu$W of incident optical power from room temperature down to 0.5 K.  

\begin{figure}[h]
\begin{center}
\includegraphics[width=1.0 \columnwidth]{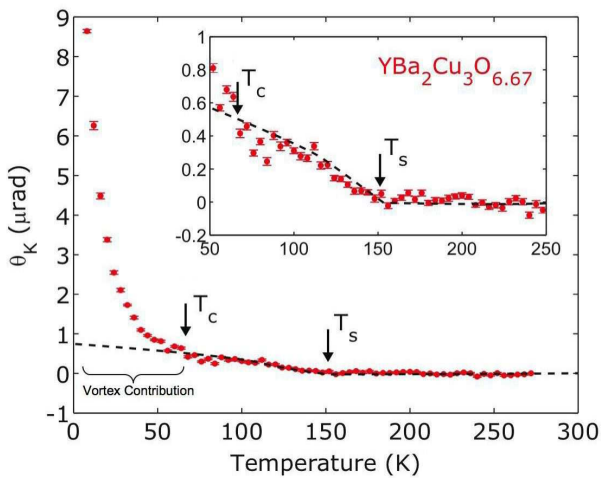}
\end{center}
\caption{ Kerr effect of YBa$_2$Cu$_3$O$_{6.67}$ crystal. The sample was first cooled to 4.2 K in a +4 T field. The field was turned off at 4.2 K and measurements were taken while warming the sample. Note the large vortex contribution which disappears just before $T_c=65K$. The inset shows the region above $T_c$ with its zero baseline, indicating a finite Kerr signal that disappears at $T_s=$155K. Dashed lines are guides to the eye.} 
\label{hf}
\end{figure}

Crystals were mounted on a copper plate using GE varnish. The system was aligned at room temperature, focusing the beam that emerges out of the quarter-waveplate to a $\sim 3 \mu$m size spot \cite{xia2}. A measurement cycle was then used in which the sample was first cooled in a field, the field was turned off at the lowest temperature (4.2 K), and the Kerr effect of the sample was measured while the sample was warmed to room temperature. Fig.~\ref{hf} shows the Kerr effect measured on YBa$_2$Cu$_3$O$_{6.67}$ after cooling the sample in a field of 4 T. Three regimes are clearly observed. The low temperature Kerr effect is very large, indicating a large contribution from trapped  vortices. This contribution, which follows the direction of the magnetic field, decays exponentially with increasing temperature and, at $T_c$, reaches a finite value that is of order $\sim 1\mu$rad.  This clearly indicates a new, unexpected state with a small but finite ferromagnetic-like signal.  As we continue to warm the sample above $T_c$, that remnant signal decreases until it disappears at a higher temperature denoted by $T_s$. Above $T_s$ the Kerr signal is zero to within our sensitivity ($\pm$ 30 nanorad). We emphasize that this is a true zero as all Kerr data shown in this paper is raw, without any baseline subtraction. Error bars in all figures are one-sigma statistical error. For the sample in Fig.~\ref{hf} we find $T_s = 155 \pm 5 K$.

\begin{figure}[h]
\begin{center}
\includegraphics[width=1.0 \columnwidth]{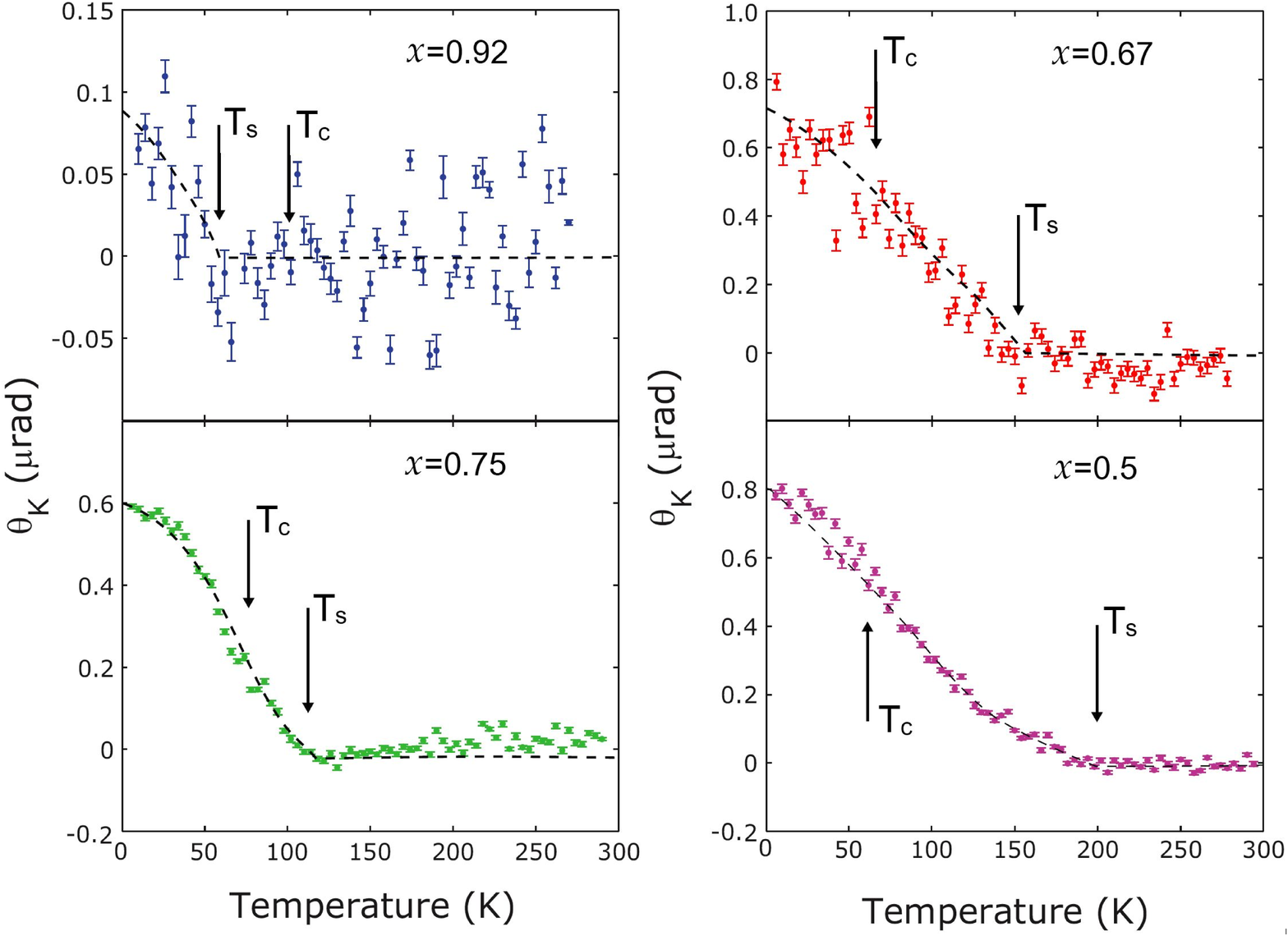}
\end{center}
\caption{ Kerr effect of zero-field warmup for $x=$ 0.92, 0.75, 0.67, and 0.5.  All samples were cooled and measured in a magnetically shielded environment (field $<$3 milli-Oe). Dashed lines are guide to the eye.} 
\label{set}
\end{figure}

One of the key features of a broken symmetry state is its sensitivity to small symmetry breaking fields.  This leads to characteristic hysteresis loops.  In particular, a ferromagnet cooled below its critical temperature in a small field will remain magnetized even when the field is turned off unless the temperature is subsequently raised above $T_c$, or an opposite field in excess of a certain temperature dependent ``coercive field" is applied to reverse the magnetization.  Therefore, to elucidate the character of the broken symmetry state detected in the present experiments, we have measured the history dependence of the Kerr rotation following several different protocols. The data in Fig.~\ref{set} were obtained on four different crystalline samples with  $x=$ 0.92, 0.75, 0.67, and 0.5  as follows:  First, the sample was ``trained" in a 4 T field in the ``up" direction at room temperature.  The field was then turned off, and the sample was cooled in zero field to 4.2K \cite{fieldcool}.  All zero field measurements were
done in conditions where all magnets were open loops at room temperature, and the system was cooled in a double $\mu$-metal-shield environment with a remnant field $<$ 3 mOe, measured separately at the position of the sample.

The Kerr signal was measured upon warming, still in zero field  (ZFC/ZFW).   It is clear that for all four samples there is a temperature $T_s$ at which a finite signal disappears  when the temperature is raised. Moreover, it is evident that while for highly underdoped samples $T_s > T_c$, near optimal doping no signal is observed above and through $T_c$ but rather appears at a temperature well below $T_c$ (see Fig.~\ref{set}a).  The crystal in Fig. 3c is the same as in Fig. 2.  Note that the two traces look essentially identical above $T_c$, while the FC/ZFW trace shows a large vortex signal below $T_c$ which is missing in the ZFC/ZFW trace. In all the underdoped samples, $T_s > T_c$.  However, it is the lack of a vortex signal in the ZFC/ZFW traces that allows us to detect  $T_s$ in the near optimally doped sample, where $T_s < T_c$. Furthermore, inspecting the temperature dependence of the Kerr signal near $T_s$, we note a ``supelinear" curvature, an observation that may point to the fact that we are probing a secondary order parameter.

\begin{figure}[h]
\begin{center}
\includegraphics[width=1.0\columnwidth]{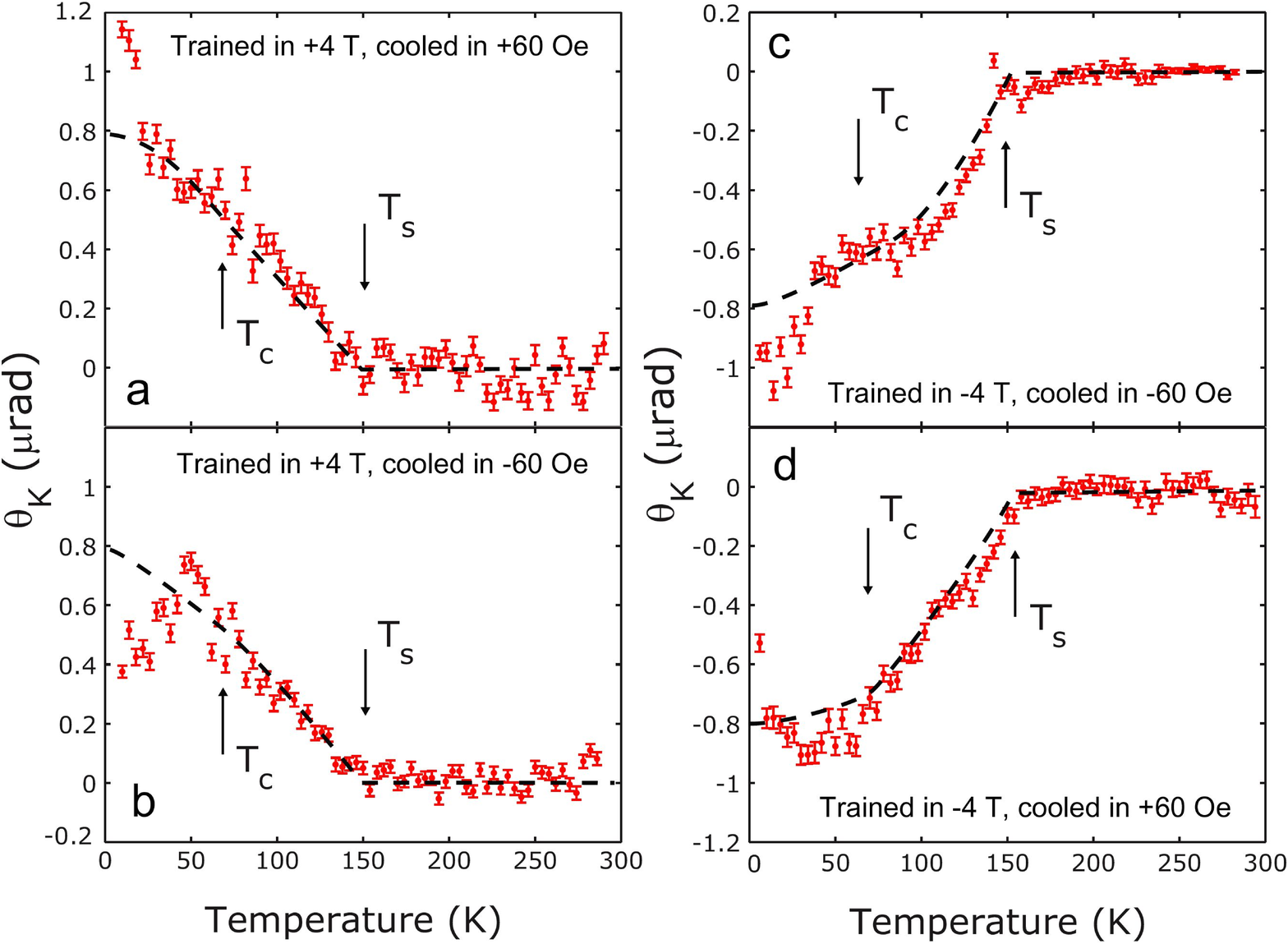}
\end{center}
\caption{ Kerr effect measurement of YBa$_2$Cu$_3$O$_{6.67}$ crystal taken while warming  the sample after cooling it in fields of -60 Oe and +60 Oe, and switching the field off at 4.2 K. These measurements were taken after the sample was trained in a field of -4 T (left) and +4 T (right), as shown in figure \ref{hf} (see text). Note the much smaller vortex contribution and the fact that it tracks the sign of the field in which it was cooled in. } 
\label{train}
\end{figure}

It is important in identifying the Kerr effect with a state with spontaneous symmetry breaking to demonstrate that it can be reversed in a sufficiently high magnetic field.  However, one very unexpected feature of our data (which is implicit already in the above) is that the coercive field does not vanish at $T_s$, nor indeed up to  room temperature.  (We have not gone above room temperature in order not to destroy the oxygen ordering in the crystals).   Both aspects of this are demonstrated by the traces shown in Fig.~\ref{train} taken on the same crystal with x=0.67 as in Figs.~\ref{hf} and Fig.~\ref{set}c.  (Similar, although not quite as extensive results have been obtained for the other crystals.)  First, the crystal is trained at room temperature in a +4 T field, which is then removed.  In Fig.~\ref{train}a, this crystal is field cooled in a "small" field of +60 Oe, and then measured in a zero-field warm-up.  Not unexpectedly, the trace looks very similar to that in Fig.~\ref{hf}, except with a substantially smaller vortex contribution below $T_c$.  In Fig.~\ref{train}b, the crystal is cooled in a small field of -60 Oe, and then measured in a zero-field warm-up.  Clearly, the vortex signal below $T_c$ has the opposite sign, as expected, but the signal above $T_c$ is unchanged by the application of the -60 Oe field, proving that the coersive field at room temperature is in excess of this value.  To obtain the traces in Figs.~\ref{train}c and \ref{train}d, the crystal was first retrained by the application of a -4 T field at room temperature, then field cooled in a field of $\pm$ 60 Oe, respectively, and measured in a zero-field warmu-up.  Clearly, these two curves are just the reverse of the curves in Figs.~\ref{train}a and \ref{train}b. A detailed study of the training hysteresis showed that a field of $\sim$3 T applied at room temperature fully  flips the orientation of the effect below $T_s$, hence the choice of 4 T for our training schedule. This study will be the subject of a future publication. 

\begin{figure}[h]
\begin{center}
\includegraphics[width=1.0\columnwidth]{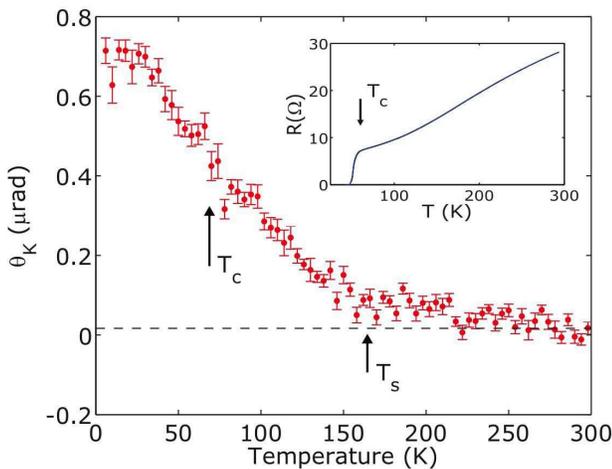}
\end{center}
\caption{Kerr effect of a c-axis film with estimated average oxygen stoichiometry of $0.5 < x <0.67$.   Inset shows the resistive transition of the sample. We also note the position of $T_c$ and the range of $T_s$ (see text) for this sample.} 
\label{film}
\end{figure}

So far we have discussed a series of single crystals of \YBCO. However, we note that all the effects we observed above were also observed on a series of thin films of underdoped \YBCO.  Such films are known to be inhomogeneous with transitions that are broader than in single crystals. This may be  due to grain boundaries, twin boundaries, disorder in the chains or  lattice  distortion induced by the substrate, among other reasons. Thus, exploring the Kerr signal in thin films will be an excellent test of the genuine origin of the effects and their sensitivity to disorder. While a detailed study of the films will be given in a different publication, we introduce Fig.~\ref{film} here as an example of a c-axis film with onset temperature $T_c \sim$60 K. Compared to the crystals, this is a sample that should be in between $x=0.5$ and $x=0.67$. Indeed the signal we observe is similar to the one observed in Fig.~\ref{set}c, with a broad onset at $T_s \sim$160 K, similar to that of the crystal with $x=0.5$ (Fig.~\ref{set}d).

The observed large training fields suggest that TRS is already broken  above room temperature, but the nature of this  state is unclear. If it has a ferromagnetic component, then its moment is less than  $\sim 10^{-5} \mu_B$/Cu which is our estimated sensitivity.  The signal that we do measure at $T_s$ cannot be a consequence of a gradual increase of the high-temperature signal because  its sharp onset resembles a true broken symmetry at $T_s$.  We therefore suggest that another order parameter orders at $T_s$ which is coupled to the high-temperature TRS-breaking order parameter.  This order parameter may either itself break TRS, or, it becomes ``visible"  through a ferromagnetic-like component that is induced in the high-temperature order through some distortion below $T_s$ .

The fact that $T_s$ marks the onset of a true symmetry breaking effect gains support from the recent elastic neutron scattering measurements  \cite{fauque,mook},  and  earlier $\mu$SR measurements \cite{sonier}.  Using polarized elastic neutron diffraction, Fauqu\'{e} {it et al.} identified a magnetic order in the \YBCO~system which does not break translational symmetry and is consistent with either opposite moments on oxygens of adjacent bonds, or two counter circulating charge current loops within the unit cell. The circulating current state was proposed by Varma \cite{varma} to account for a symmetry breaking effect at $T^*$.  We note that the first possibility necessarily produces a small ferromagnetic moment due to the orthorhombicity of YBCO.  The current loop state, by itself, is incompatible with  ferromagnetism, but a ferromagnetic component can be induced by any additional effect (e.g. impurities), which further reduces the spatial (point-group) symmetry \cite{aji} of the crystal. Comparing  the neutron data to our $T_s$ reveals that the onset of the effect for similar dopings is  $\sim$ 30  K higher for the neutron experiments.  The $\mu$SR measurements were done for $x=$0.67 and for $x=$0.95, both give onset temperatures for increased muon relaxation that are identical to our $T_s$. Both, the neutron and $\mu$SR experiments  also note the existence of magnetic scattering above the onset temperature that persists all the way to room temperature. However, no proposal has been put forward to explain this effect.

In conclusion we have reported in this paper the discovery of a novel magnetic order in a wide range of doping of \YBCO. The new effect is ferromagnetic-like and onsets at a temperature that matches 
the pseudogap behavior in underdoped cuprates. We further find evidence that the line defined by the onset of this effect crosses the superconducting dome to appear below $T_c$ for near optimally doped sample.  Finally we find that this effect couples to another time reversal symmetry effect that occurs at high temperatures and dictates the sign of the Kerr effect that appears at the pseudogap temperature.

\acknowledgments
Discussions with D. Fisher, D. Scalapino,  and especially C. Varma  are greatly appreciated.  This work was supported by Stanford's CPN (NSF NSEC Grant 0425897) and by DoE   grant DEFG03-01ER45925.


\begin{thebibliography}{99}
\bibitem{fieldcool}
Alternatively the sample was cooled in an applied field of 4 T to 4.2 K at which the field was turned to zero. The sample was then warmed  to room temperature and placed into the magnetically shielded environment for ZFC/ZFW measurements. Both high field training procedures yielded the same Kerr result shown in Fig.~\ref{set}.
\end{thebibliography}
\end{document}